On "Nanoscale thermodynamics needs the concept of a disjoining chemical potential", by W. Dong, *Nat. Comm.* **10**, 1038 (2023).

Ralph V. Chamberlin; Department of Physics, Arizona State University, Tempe, AZ USA 85287-1504

In a recent Article [1], W. Dong argues that a disjoining chemical potential is needed to explain finite-size effects in thermodynamics. Dong then claims that this chemical potential should be used instead of Hill's nanothermodynamics because Hill's subdivision potential "has never been measured experimentally" and "remains elusive to relate…to any experimental observables." Here, I challenge this claim by reviewing some of the experimental evidence for Hill's subdivision potential found in most types of materials [2-6], including the corrections to scaling that are measured near ferromagnetic transitions and explained by a mean-field cluster model in a Letter to Nature from 2000 [7] where the term "nanothermodynamics" first appeared [8]. Because other uses of this term now occur [9], my primary goals are to: clarify how Hill's work was extended in [7] to explain many measurements, propose ways to reduce the confusion conveyed by [1], and how the predictions in [1] can be improved for real systems by adding nanothermodynamics.

Thermal and dynamic heterogeneity inside bulk samples are experimental observables that can be related to the subdivision potential in nanothermodynamics. In general, such heterogeneity occurs in thermal equilibrium only in the nanocanonical ensemble of nanothermodynamics. Specifically, in standard thermodynamics all systems must be effectively infinite and homogeneous, whereas several experimental techniques have shown that heterogeneity is ubiquitous in most types of materials, including liquids and glasses [2-4], spin glasses [10], polymers [11], and crystals [7,5]. Even for the semi-classical ideal gas, heterogeneity from nanothermodynamics provides a novel solution to Gibbs' paradox [6], with the statistics of indistinguishable particles needed only on microscopic length scales, so that particles can be distinguished by their location when separated macroscopically. For all such systems that subdivide into internal subsystems, the theoretical "replica trick" from small-system thermodynamics that is purposely avoided in [1], becomes an unavoidable physical reality in nanothermodynamics.

A brief history of the term "nanothermodynamics" may help explain its misuse in [1]. Hill conceived of the term as a new name for the theory of small-system thermodynamics that he developed in 1961-63 [8]. However, the term was first used in 1999 during discussions of how concepts from small-system thermodynamics can be extended to large systems that subdivide into small subsystems [12], leading to the first published use of the term in 2000 [7]. If these subsystems are statistically independent they form an ensemble of subsystems inside a bulk sample, which is conceptually distinct from (but mathematically similar to) Hill's theory of a large ensemble built up from small systems. Now, two different interpretations of "nanothermodynamics" appear in the literature. In [5,6], the difference between the two pictures is clarified, with the suggestion that if only one aspect is to be discussed then different names should be used for the distinct interpretations, specifically small-system thermodynamics for Hill's original picture and nanothermodynamics for small subsystems inside bulk samples. Then, [1] and [7] can be reconciled by: Dong considers only small-system thermodynamics, with no discussion of (or reference to) the experimental evidence and conceptual implications from nanothermodynamics.

In [1] it is implied that the main reason for considering a disjoining chemical potential is its relevance to experiments. Although the theory and simulations in [1] are based on an ideal gas, Dong argues that a similar disjoining chemical potential should apply to real systems. However, simulations in [1] will fail to match the measured thermal behavior of any real gas that obeys the Sackur-Tetrode equation, which is well established in the laboratory [13]. Thus, for measurements on many real substances, the predictions in [1] can be improved by adding nanothermodynamics to facilitate realistic statistics of indistinguishable particles on the scale of nanometers [6].

Figure 1 shows one way to picture the importance of Hill's fundamental equation of thermodynamics (combined first and second laws) that includes finite-size effects [6]. In Fig. 1, Hill's equation is mapped to a simple (three-energy-level) system for distinct contributions to conservation of energy during reversible changes. The first three terms on the right side of the equation come from Gibbs' fundamental equation. The fourth term contains Hill's subdivision potential ($\mathcal{E}$) and number of subdivisions ($\eta$), which are useful for small systems, and essential for subsystems inside bulk materials. The diagram on the left side of the equation represents an initial distribution of energy states in the three levels. The first diagram on the right side of the equation shows how adding heat changes the distribution of energies, without changing the levels. The second diagram on the right shows how work done on the system changes the levels, without changing the distribution. The penultimate diagram shows how adding particles changes the total occupation of states, without changing their distribution. The final diagram shows how changing the size of the small system (or subsystem) changes the widths of the energy levels due to changing surface states, length-scale terms, thermal fluctuations, etc. Standard thermodynamics, which requires systems to be effectively infinite and homogeneous, has no systematic way of including all of these contributions. For example, the fully-open generalized ensemble (first named the nanocanonical ensemble in 2006 [14]) is ill-defined in standard thermodynamics. Because none of the extensive environmental variables is fixed in this ensemble, their fluctuations are anomalously large. For example, energies and sizes of small systems have mean-squared fluctuations that increase with the square of the number of particles, not linearly as in standard thermodynamics [15]. Such large fluctuations can only be evaluated without artificial restraints in the generalized ensemble of Hill's small-system thermodynamics, or the nanocanonical ensemble for nanothermodynamics of small subsystems inside large systems.

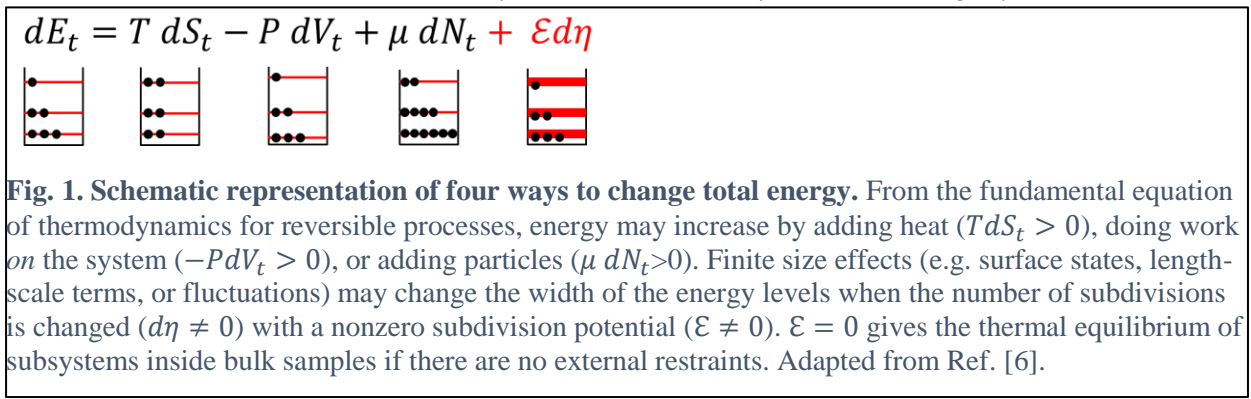

**Fig. 1. Schematic representation of four ways to change total energy.** From the fundamental equation of thermodynamics for reversible processes, energy may increase by adding heat ($TdS_t > 0$), doing work on the system ($-PdV_t > 0$), or adding particles ($\mu\, dN_t > 0$). Finite size effects (e.g. surface states, length-scale terms, or fluctuations) may change the width of the energy levels when the number of subdivisions is changed ($d\eta \neq 0$) with a nonzero subdivision potential ($\mathcal{E} \neq 0$). $\mathcal{E} = 0$ gives the thermal equilibrium of subsystems inside bulk samples if there are no external restraints. Adapted from Ref. [6].

In many applications, including those for the measured behavior of ferromagnetic materials in [7], the subdivision potential is set to zero, $\mathcal{E} = 0$. Note that $\mathcal{E} = 0$ is NOT the same as a large system with no subdivisions, $\mathcal{E} \to 0$. How and why $\mathcal{E} = 0$ is needed for stability in the generalized ensemble is discussed at length in Section 10-3 of [16]. Indeed, from Pg. 101: "Although $\mathcal{E}$ is negligible for a macroscopic system (…), it is *not* equal to zero in the strict sense that we are using $\mathcal{E} = 0$ above as an equilibrium condition. The macroscopic state is therefore not to be confused with the equilibrium state." An analogy is when the chemical potential is set to zero ($\mu = 0$) to ensure that artificial restraints do not alter the distribution of some types of particles, e.g. photons and phonons. Similarly, $\mathcal{E} = 0$ ensures that artificial restraints do not alter the average size and distribution of subsystems. In general, $\mathcal{E} \neq 0$ while a system changes its distribution of subsystems, until it reaches $\mathcal{E} = 0$ where the total free energy of the system is minimized (Eq. 10-116 in [16]), yielding the thermal equilibrium nanocanonical ensemble. Although Hill often considered finite-size thermal effects under diverse conditions, by 1964 he already knew that the equilibrium state requires $\mathcal{E} = 0$. Because this concept of $\mathcal{E} = 0$ was first applied to bulk samples in 1999 [17] and named "nanothermodynamics" in 2000 [7], it could be renamed stable nanothermodynamics. Then, the term "Hill's nanothermodynamics" used by many authors that focus on small systems with $\mathcal{E} \neq$

0 (e.g. [1] and [9]), without referencing the experimental evidence and conceptual implications of $\mathcal{E} = 0$, could reduce confusion by using the name out-of-equilibrium, frozen, or nonstable nanothermodynamics. However, I recommend restoring the term small-system thermodynamics, especially for small systems that have $\mathcal{E} \neq 0$, thus preserving the historical record and legacy of Hill whose seminal work 60 years ago already recognized that $\mathcal{E} = 0$ is needed for stable equilibrium in the generalized ensemble.

Nanocanonical behavior of internal subsystems can be calculated analytically for simple systems, such as the semi-classical ideal gas and one-dimensional Ising model [6]. One mechanism driving subdivision is a net increase in entropy for the nanocanonical ensemble. Figure 2 is a cartoon sketch showing how the total entropy of a simple system containing two semi-classical ideal gas particles (top) decreases when subdivided into canonical subsystems (middle), but increases if the subsystems form a nanocanonical ensemble (bottom). Here, the "walls" between subsystems can be thought of as dynamical boundaries that break quantum entanglement, so that particles in one subsystem are distinguishable from particles in other subsystems. Thus, this nanocanonical ensemble yields a novel solution to Gibbs' paradox that allows semi-classical ideal gas particles to be distinguishable by their location when separated by macroscopic distances, and sub-additive entropies that are required for quantum mechanics [18].

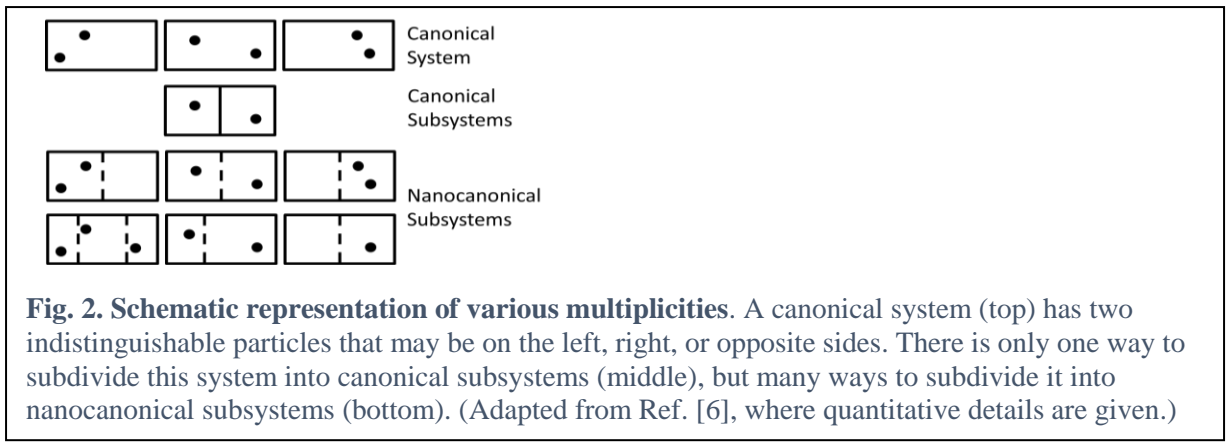

**Fig. 2. Schematic representation of various multiplicities**. A canonical system (top) has two indistinguishable particles that may be on the left, right, or opposite sides. There is only one way to subdivide this system into canonical subsystems (middle), but many ways to subdivide it into nanocanonical subsystems (bottom). (Adapted from Ref. [6], where quantitative details are given.)

As in [1], many recent studies of small-system thermodynamics [9] have focused on explaining computer simulations, where various non-stable ensembles of small systems are readily created. In contrast, most studies of nanothermodynamics have focused on explaining laboratory experiments [7,5,6], especially measurements on samples that exhibit evidence for stable ensembles of internal subsystems. In this case, the assumption of a homogeneous macroscopic system from standard thermodynamics ($\mathcal{E} \to 0$) is replaced by stable nanothermodynamics ($\mathcal{E} = 0$), yielding the nanocanonical ensemble of independent internal subsystems consistent with the thermal and dynamic heterogeneity measured in most types of materials [2-7,10,11]. In fact, quantitative agreement is found between models based on nanothermodynamics and the measured temperature dependence of the magnetic correlation range ($\xi$) in cobalt [7], structural correlations in $LaMnO_3$ [5], and the sizes of independently relaxing regions in glass-forming liquids [19]. Furthermore, a mean-field Ising model that includes finite-size effects from the nanocanonical ensemble yields improved agreement with measured corrections to scaling near ferromagnetic phase transitions by maintaining $\mathcal{E} = 0$ as $\xi \to \infty$ (solid symbols in Fig. 1 of [7]). Essentially the same model yields a gradual glass transition that is smeared out by finite-size effects when the correlation range is limited by disorder [17]. Moreover, nanothermodynamics extends the 2[nd] law of thermodynamics to nanoscale subsystems, providing improved agreement with measured 1/$f$-like noise from qubits [6], metal films, spin glasses, and nanopores [20]. To summarize, the "replica trick" from small-system thermodynamics that is purposely avoided in [1] becomes an unavoidable physical reality in the stable nanothermodynamics used to explain many diverse measurements and theoretical expectations.